\documentclass[aps,10pt,pra,twocolumn,floatfix,floats,showpacs,superscriptaddress,raggedbottom]{revtex4-1}
\usepackage{graphicx,latexsym}
\usepackage{dcolumn}
\usepackage{amsmath}
\usepackage{amssymb,bm}
\usepackage{braket}
\usepackage{natbib}

\usepackage[hidelinks=true,breaklinks=true]{hyperref}
\hypersetup{
  colorlinks   = true, 
  urlcolor     = blue, 
  linkcolor    = blue, 
  citecolor    = red 
}

\usepackage[normalem]{ulem}

\def\sec#1{Sec.\ \ref{#1}}
\def\eq#1{Eq.\ (\ref{#1})}
\def\fig#1{Fig.\ \ref{#1}}

\begin{document}

\title{Manifestation of the Purcell effect in\\ current transport through a dot-cavity-QED system}

\author{Nzar Rauf Abdullah}
\email{nzar.r.abdullah@gmail.com}
\affiliation{Physics Department, College of Science, 
             University of Sulaimani, Kurdistan Region, Iraq}
\affiliation{Komar Research Center, Komar University of Science and Technology, Sulaimani, Iraq}

\author{Chi-Shung Tang}
\affiliation{Department of Mechanical Engineering,
  National United University, 1, Lienda, Miaoli 36003, Taiwan}

\author{Andrei Manolescu}
\affiliation{Reykjavik University, School of Science and Engineering,
              Menntavegur 1, IS-101 Reykjavik, Iceland}

\author{Vidar Gudmundsson}
\email{vidar@hi.is}
 \affiliation{Science Institute, University of Iceland,
        Dunhaga 3, IS-107 Reykjavik, Iceland}

%

\begin{abstract}
We study the transport properties of a wire-dot system coupled to a cavity and a photon reservoir. Tuning the photon energy, 
Rabi-resonant states emerge and in turn resonant current peaks are observed. We demonstrate the effects of the 
cavity-photon reservoir coupling, the mean photon number in the reservoir, the electron-photon coupling and 
the photon polarization on the intraband transitions occurring between the Rabi-resonant states, and on the 
corresponding resonant current peaks.
The Rabi-splitting can be controlled by the photon polarization and the electron-photon coupling strength.
In the selected range of parameters, we observe the results of the Purcell effect enhancing the current peaks 
through the cavity by increasing the cavity-reservoir coupling, while they decrease with increasing the electron-photon coupling. 
In addition, the resonant current peaks are also sensitive to the mean number of
photons in the reservoir.
\end{abstract}



\maketitle

%
%

\section{Introduction}\label{Sec:Introduction}

Single photon sources have been widely sought after for research in fields of science and 
technology \cite{Imamog72.210,Giannelli_2018}.
A single photon can be used to control optical quantum simulators \cite{Aspuru-Guzik2012, doi:10.1063/1.4977023} and 
multi-qubit gates \cite{Lin2015},
and it can also be coupled to an electronic structure such as a quantum dot (QD) to control electron motion \cite{Hanschke2018}.
In such systems, quantum mechanical methods are used to describe the light consisting of few photons, i.e., 
the light field has to be fully quantized \cite{Giannelli_2018, Kreinberg2018}. 
A quantized photon system coupled to an electronic system can be used to explore many interesting aspect of physical 
problems and phenomena in the nanoscale range such as the Purcell effect \cite{doi:10.1021/nl3008083}, quantum information processing \cite{FriskKockum2019}, 
quantum communication networks \cite{DeGreve2012} and other applications of 
nanotechnology \cite{PhysRevB.87.115419,PhysRevB.91.205417,Delbecq2013,PhysRevB.78.125308}.

Several parameters need to be considered when studying light-matter interactions such as 
electron-photon coupling strength,  $g_\gamma$, \cite{PhysRevB.91.205417}, the coupling strength of the cavity-photon field
to the environment, $\kappa$,~\cite{doi:10.1063/1.3294298,DELVALLE2011241} and mean value of photons in the environment 
with energy corresponding to the cavity mode, i.\ e.\ the temperature of the environment (which is the photon reservoir). 
The electron-photon coupling strength can be compared to the coupling strength of the cavity-photon field
to the environment. If the electron-photon coupling strength is greater than the coupling strength of the cavity photon
to the environment, $g_{\gamma} > \kappa$, the system is said to be in the strong coupling regime~\cite{PhysRevLett.121.043601}.

In the strong coupling regime, a QD system exposed to a quantized photon field has been found to be one of the 
most fascinating system for investigating several physical phenomena in modern nanodevices. The Rabi splitting and oscillations 
in a QD coupled to a photon field lead to observation current peaks which can be used to measure photoluminescence~\cite{Leng2018}, 
vacuum effects, like the ground state electroluminescence \cite{PhysRevLett.116.113601,doi:10.1002/andp.201700334},
entanglement characteristics of a photon source with an electronic system~\cite{PhysRevLett.109.240501}, 
resonance fluorescence and Rayleigh scattering in Mollow-triplet-like spectra~\cite{PhysRevLett.106.243601}, and the
photon-induced transport in a Rabi-splitting of a two level~\cite{Faraon2008} and many level QD~\cite{Vidar-ACS-Phot}.
Furthermore, the strong coupling regime paves the way to industrial technology for building solid state-based quantum  optical processors~\cite{Wu2019} and semiconductor chips~\cite{Sillanpaa2007, Majer2007}.

An early interesting achievement in the field of quantum optics was the demonstration of 
the Purcell effect \cite{PhysRev.69.37} which is the the enhancement of a quantum system's spontaneous 
emission rate by its environment.
This phenomena has been investigated by many research group \cite{Kiraz_2003,PhysRevLett.120.114301}.
The emission intensity of spin-up exciton state with respect to spin-down exciton state is enhanced at resonance 
due to Purcell effect \cite{doi:10.1021/nl3008083}. In addition, the enhancement of the Purcell effect has been achieved 
by controlling the effective coupling with the microcavity \cite{PhysRevLett.98.063601,PhysRevLett.98.117402}.

Motivated by the above mentioned scientific works, we model a two-dimensional electron system in a 
GaAs QD embedded in a short quantum wire coupled to two electron reservoirs \cite{Zhang2016}. 
The wire-dot system is also coupled to a 3D-cavity and the cavity is in turn coupled to a 
photon reservoir, i.\ e.\ the external environment. The electron transport in the steady-state is investigated under 
the effects of a cavity photon field using a Markovian quantum master equation~\cite{JONSSON201781}. 
Previously, we have investigated and reported the characteristics of electron transport in systems with different geometries 
using non-Markovian for the short time evolution, and Markovian master equations for the long time, 
including the coupling of electrons to a quantized photon field.
We have studied Rabi-oscillations in the intermediate time regime~\cite{GUDMUNDSSON_2019}, oscillations in the electron 
transport caused by Rabi-resonant states in the steady state~\cite{Nzar-arXiv_article_2019}, and photon-assisted
tunneling~\cite{PhysicaE.64.254} and thermoelectric transport in the transient 
regime~\cite{Nzar.25.465302, Nzar_ACS2016,nzar27.015301,Vidar85.075306,Abdullah2017}. 
In addition, the photocurrent generated by photon replica states in an off-resonant dot-cavity system has been presented 
where the Rabi effect plays only a minor role in the transport because the system is in a off-resonant regime. 
It has been shown that the photocurrent can be manipulated by the photon polarization 
and the cavity-photon coupling strength of the environment~\cite{Nzar-arXiv_article_2019_1}. 
Furthermore, the resonant current peaks generated by Rabi-resonant states in a quantum dot have been 
investigated where only the influences of the photon polarization was highlighted. We have shown that the Rabi effect 
has a major impact on the transport~\cite{Nzar-arXiv_article_2019}.
In the current work, we present a general picture of the influences of the electron-photon coupling strength, 
the cavity-reservoir coupling strength and the mean photon number in the photon reservoir on the transport 
properties of a QD system in the steady-state regime in which the Rabi-effect has a large role.
We assume a strong coupling regime, ($g_{\gamma} > \kappa$), and investigate the resonant current generated by the multiple 
resonance states. An enhancement of the current through the QD system is observed, which is 
demonstrated to be a direct consequence of the Purcell effect.
The rest of the paper is presented as follows. We define the model system in~\sec{Sec:Model}.
Results are discussed for the model in~\sec{Sec:Results}. Finally, we show our conclusion in~\sec{Sec:Conclusion}.

\section{Hamiltonian of the total system}\label{Sec:Model}

We assume a QD embedded in a two dimensional short quantum wire in the $xy$-plane with 
hard walls at the ends in the $x$-direction and the parabolic confinement potential in the $y$-direction.
The wire-QD system is exposed to a constant external weak magnetic field and 
coupled to a photon cavity with a single photon mode.
In order to pump electrons to and from the QD system, we assume the QD system coupled to 
two electron reservoirs via a tunneling region called the coupling or contact region~\cite{Vidar61.305,GUDMUNDSSON20181672}.
The Hamiltonian of the total system, the QD system and the cavity,
in the many-body basis can be defined as 
\begin{equation}
 \hat{H}_{\rm S} = \hat{H}_e + \hat{H}_{\gamma} + \hat{H}_{e\text{-}\gamma},
 \label{H_S}
\end{equation}
where $\hat{H}_e$ is the Hamiltonian of the QD system,  $\hat{H}_{\gamma}$ indicates the Hamiltonian of the free photon field, and 
$\hat{H}_{e\text{-}\gamma}$ defines the interaction between the QD system and the cavity. 
We start with the Hamiltonian of the QD system which is given by
\begin{align}
  \hat{H}_e & = \sum \limits_{nn'}^{} \bra{\Psi_{n'}} \Big[ \frac{\hat{\bm \pi}_e}{2m_{\rm eff}} + e V_{\rm g} + V_{\rm QD} \Big] \ket{\Psi_n} d^{\dagger}_{n'} d_n  \nonumber \\
              & + H_{\rm Z} + \frac{1}{2}\sum_{\substack{nn' \\ mm'}} V_{nn'mm'} \, d^{\dagger}_{n'} d^{\dagger}_{m'} d_m  d_n.
              \label{H_e}
\end{align}
Herein, $\Psi_{n}$ stands for a single electron eigenstate, $m_{\rm eff}$ is the effective mass of electrons, and 
$\hat{{\bm \pi}}_e =: {\bf p}+\frac{e}{c} {\bf A}_{\rm B}$, where ${\bf p}$ is the canonical momentum operator, 
$\mathbf{A}_{\rm B} = -By \mathbf{\hat{x}}$ is the magnetic vector potential with $\mathbf{B} = B \mathbf{\hat{z}}$.
$V_{\rm g}$ is the gate voltage that shifts the energy states of the QD system with respect to the chemical potentials of 
the leads, $V_{\rm QD}$ is the potential that forms the quantum dot,
and $d^{\dagger} (d)$ are the fermionic creation (annihilation) operators.

The first term of the second line of \eq{H_e} is the Zeeman Hamiltonian,  $H_{\rm Z}= \pm g^{*}\mu_B B  \sigma_z/2$, and
the second term represents the Coulomb interaction, while $V_{nn'mm'}$ are the Coulomb integrals
\begin{equation}
V_{nn'mm'} = \bra{\Psi_{n'} \Psi_{m'}} \frac{e^2}{\bar{\kappa} |r - r'|} \ket{\Psi_n \Psi_m},
\end{equation}
where $\bar{\kappa}$ is the dielectric constant, and $|r - r'|$ 
the spatial separation of an electron pair.
An exact diagonalization technique in a truncated Fock space is used to obtain 
the many-electron states.

The second term of \eq{H_S} defines the free photon field 
\begin{equation}
 \hat{H}_{\gamma} =  \hbar \omega_{\gamma} \, a^{\dagger} a
\end{equation}
with $\hbar \omega_{\gamma}$ being the single photon energy 
and $a^{\dagger}$ and $a$ the bosonic creation and annihilation operators, respectively.
The last term of \eq{H_S} is 
\begin{align}
 \hat{H}_{e\text{-}\gamma} & = \frac{e}{m_{\rm eff}}  \sum_{nn'}  \bra{\Psi_{n'}} {\bm \pi_e} \cdot {\bm A_{\gamma}} \ket{\Psi_{n}} \, d^{\dagger}_{n'} d_n \nonumber \\
         & + \frac{e^2 {\bm A_{\gamma}}^2}{2 m_{\rm eff}} \sum_{nn'} \langle \Psi_{n'} | \Psi_{n} \rangle \, d^{\dagger}_{n'} d_n,
         \label{H_e_ph}
\end{align}
which is the Hamiltonian of the electron-photon interactions including both the
paramagnetic Hamiltonian ($A_{\gamma}$ term) and the diamagnetic Hamiltonian ($A_{\gamma}^2$ term). 
The photon field interacting with the QD system is represented via the vector potential 
\begin{equation}
 {\bf A}_{\gamma} = A({\bf e} \, a + {\bf e}^*  \, a^{\dagger})
\end{equation}
with ${\bf e} = {\bf e}_x$ for $x$-polarized photon field and 
${\bf e} = {\bf e}_y$ for $y$-polarized photon field.
The amplitude of the photon field, $A$, is related to the electron-photon coupling strength via 
$g_{\gamma} = e A a_w \Omega_w/c$, where $a_w$ is the effective magnetic length, $e$ displays the electron charge, 
and $\Omega_w$ refers the effective confinement frequency of electrons in the QD system. We assume the 
wavelength of the cavity field to be much larger than the size of the electronic system
composed of the short wire and the quantum dot. A numerically exact diagonalization procedure 
is used for the electron-photon interaction using the Coulomb interacting many-body bases
obtained earlier \cite{Vidar61.305}.

The QD system is coupled to two leads from the left and right sides, with 
different chemical potential. Therefore, electrons can flow from the leads to the QD system, and vice versa. 
To calculate the electron motion through the system in the steady-state regime, 
a Markovian quantum master equation is utilized. 
The derivation of the master equation formalism starts with the projection formalism of Nakajima 
and Zwanzing~\cite{Zwanzing.33.1338,Nakajima20.948}. The resulting non-Markovian generalized master
equation (GME) with an integral kernel evaluated up to second order in the system-lead coupling delivers
the reduced density operator for the central system.  As we are interested in the steady-state
we apply a Markovian approximation to the GME and transform it to Liouville space of 
transitions~\cite{JONSSON201781}. 
One assumes the initial reduced density operator of the QD system to be $\hat{\rho}_{\textrm{S}}(t_0)$ 
and for the leads it is $\hat{\rho}_{l}(t_0)$. Before the coupling between the QD system and the leads, the total 
density operator is assumed to be a tensor product of the uncorrelated sub parts $\hat{\rho}(t_0) = \hat{\rho}_{l}(t_0) \hat{\rho}_{\textrm{S}}(t_0)$.
After the the coupling, for $t > t_0$, we can write the reduced density operator of the QD system as 
$\hat{\rho}_{\textrm{S}}(t) = \textrm{Tr}_{l}(\hat{\rho})$ with $l$ expressing
the left (L), the right (R) leads and the photon reservoir. 
For the electron-photon interaction in the central system we do not use the rotating wave approximation, 
but we do so for the photon-cavity environment coupling. In order to do that properly we have taken care to
rid the annihilation(creation) operator for the cavity photons in the dissipation terms of the master equation
of all high frequency creation(annihilation) terms when casting the master equation into the fully interacting
basis of cavity-photon dressed many-electron states \cite{PhysRev.129.2342,PhysRevA.31.3761,PhysRevA.84.043832}.

Once we obtain the reduced density operator of the QD system, the current going through 
it can be calculated using 
\begin{equation}\label{Eq_2}
 I_{L,R} =: \mathrm{Tr}[\hat{\dot{\rho}}_S^{L,R}(t) \hat{Q}] .
\end{equation}
where $Q = -e \sum_i d_i^\dagger d_i$ is the charge operator of the QD system 
with $\hat{d}^\dagger (\hat{d})$ the electron creation (annihilation) operator of the QD system, respectively~\cite{Vidar:ANDP201500298}. 

\section{Results}\label{Sec:Results}

The main results of our calculations are presented in this section.
We consider the diameter of the quantum dot to be $d = 66.5$~nm and the length of the quantum wire to be 
$L_x = 150$~nm. A weak external magnetic field is applied to the total system, the QD system and the leads, 
$B = 0.1$~T, which is perpendicular to the two-dimensional plane of the electron motion. 
The magnetic field is weak enough to avoid most of the influences of Lorentz force on the orbital motion
and lifts the spin degeneracy of the system. 
The chemical potentials are assumed to be $\mu_L = 1.65$~meV and $\mu_R = 1.55$~meV here, and the temperature of 
the leads is fixed at $T_{\rm L,R} = 0.5$~K. 

\begin{figure}[htb]
\centering
    \includegraphics[width=0.5\textwidth,angle=0]{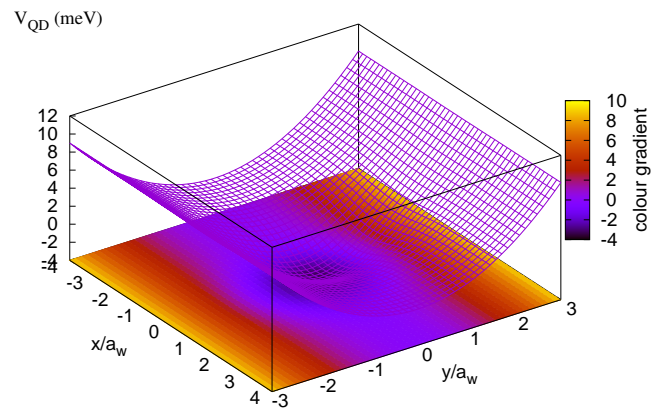}
 \caption{The potential forms the QD embedded 
 in a quantum wire that will be coupled diametrically to the left and right leads
 in the $x$-direction}
\label{fig01}
\end{figure}

We intend to show the influence of tuning the photon energy, $\hbar \omega_{\gamma}$, the photon polarization, 
the cavity-reservoir coupling strength, $\kappa$, and the mean photon number in the reservoir, $n_R$, on the transport 
properties of the QD system.
The two physical parameters, $\kappa$ and $n_R$, are included in the Markovian master equation which is not presented here~\cite{GUDMUNDSSON20181672}.
Figure \ref{fig01} shows the QD represented by the potential
\begin{equation}
  V_{\rm QD} = V_0 \, e^{(-\gamma_x^2 x^2 - \gamma_y^2 y^2)},
\end{equation}
embedded in a quantum wire. $V_0$ is the depth of the QD, 
$\gamma_x$ and $\gamma_y$ together with $V_0$ determine the diameter of the QD. 
We assume $V_0 = -3.3$~meV and $\gamma_x = \gamma_y = 0.03$~nm$^{-1}$ in our calculations.

\begin{figure}[htb]
\centering
    \includegraphics[width=0.23\textwidth,angle=0]{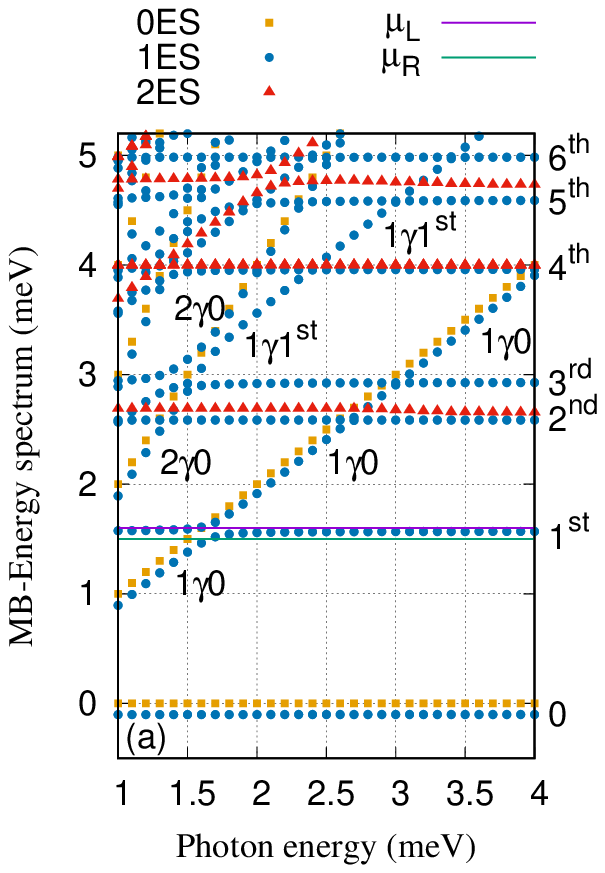}
    \includegraphics[width=0.23\textwidth,angle=0]{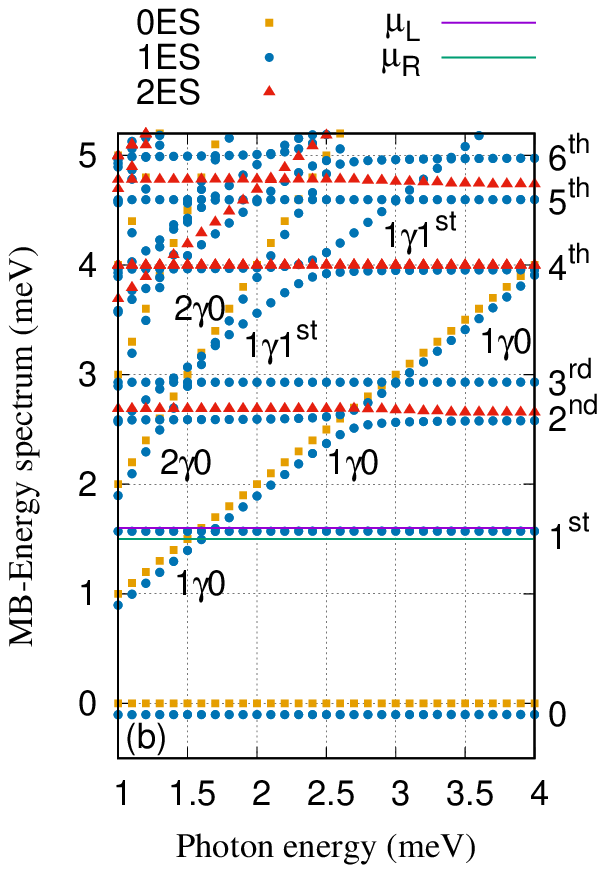}
 \caption{Many-Body energy spectra of the QD system coupled to the photon field as a function of 
        the photon energy for $x$- (a) and $y$-polarized (b) photon field,
        where 0ES (brown squares) are zero-electron states, 1ES (blue circles) are one-electron states, 
        and 2ES (red triangles) are two-electron states.
        The chemical potential of the left lead is $\mu_L = 1.65$~meV (purple line) and the right lead is $\mu_R = 1.55$~meV (green line).
        0 indicates the one-electron ground-state energy, $1\gamma$0 and $2\gamma$0 refer to the one- and two-photon replica
        of the 0, and 1$^{\rm st}$, 2$^{\rm nd}$, 3$^{\rm rd}$, 4$^{\rm th}$, 5$^{\rm th}$, 6$^{\rm th}$ display the one-electron first-, second-, third-, fourth-, fifth- and sixth-excited state, 
        respectively. The $1\gamma$1$^{\rm st}$ indicates the one-photon replica state of the 1$^{\rm st}$.
        The electron-photon coupling strength $g_{\gamma} = 0.1$~meV, the cavity-reservoir coupling  $\kappa = 10^{-5}$ meV, and 
        the photon number initially in the cavity $n_\mathrm{R} = 1$.
        The magnetic field is $B = 0.1~{\rm T}$, $eV_{\rm g} = 0.651$~meV, $T_{\rm L, R} = 0.5$~K and $\hbar \Omega_0 = 2.0~{\rm meV}$.}
\label{fig02}
\end{figure}

The energy spectrum of the QD system coupled to the cavity in the $x$- (a) and $y$-polarized (b) 
photon field is displayed in \fig{fig02}, where the electron-photon coupling strength is $g_{\gamma} = 0.1$~meV, 
the cavity-reservoir coupling strength is $\kappa = 10^{-5}$ meV and the mean photon number in the reservoir is $n_R = 1$. 
The six lowest states of the QD system are found in the selected range of the energy between $-1$~meV and $5.2$~meV. 
The states of the system are classified as follows:
The zero-electron states, 0ES (brown squares), the one-electron states, 1ES (blue circles),  
and the two-electron states, 2ES (red triangles). 
In addition, the labels appearing in the figures, 0, 1$\gamma$0, and 2$\gamma$0 refer to 
the ground-state, the first and second photon replica of the ground-state, respectively, while 
1$^{\rm st}$ and 1$\gamma$1$^{\rm st}$ are the first-excited state and the first photon 
replica of first-excited states, respectively. 
The photon dressed many-electron states are photon replica states which have a mean number of photons 
close to integers if they are not states in a Rabi-split pair.
The rest of labels such as 2$^{\rm nd}$, 3$^{\rm rd}$, 4$^{\rm th}$, 5$^{\rm th}$, and 6$^{\rm th}$ 
indicate the second-, third-, fourth-, fifth-, and sixth-excited states, respectively. 
We note that each electronic state in the energy spectrum contains a spin 
components that are Zeeman split due to the external field $B = 0.1$~T. 
The chemical potentials of the leads (purple and green lines) are 
arranged in the way that the first-excited state, 1$^{\rm st}$, is located in the bias window.
By changing the photon energy, anti-crossings between the energy states are formed especially at 
photon energies $1.5$, $1.7$, $2.7$, $3.0$, and $3.4$~meV. The photon-exchange between the two states 
forming the anti-crossings confirms Rabi-splittings~\cite{Nzar-arXiv_article_2019}. 
It is clearly seen that the Rabi-splittings is influenced by 
the photon polarization. Therefore, the energy splitting between 1$\gamma$0 and $1^{\rm st}$ at the photon energy 
$1.7$~meV for the $x$-polarization is larger than that of the $y$-polarization. Contrary, 
the energy splitting between 1$\gamma$0 and $2^{\rm nd}$ at the photon energy $2.7$~meV for the $y$-polarization 
is larger than that of the $x$-polarization. 
This indicates that the geometry of the states plays an important role as some states are more polarizable in 
the $x$-direction and some other states in the $y$-direction. 

The influence of photon polarization on the current in the QD system was reported in 
\cite{Nzar-arXiv_article_2019}.
Tuning the photon energy several peaks in the current were observed indicating resonances.
The main peaks are found at the photon energy $1.7$~meV representing a transition between 1${\gamma}$0 and 1$^{\rm st}$,  
at $2.7$~meV for transition between 
1${\gamma}$0 and 2$^{\rm nd}$, and at $3.4$~meV for transition between 1${\gamma}$1$^{\rm st}$ and 6$^{\rm th}$ 
corresponding to the Rabi-splittings shown in \fig{fig02}.
We observed that the broadening of the current peaks depends on the strength of the corresponding Rabi-splitting \cite{Kim2015}. 
Therefore, the broadening of the current peak corresponding to the resonance between 1${\gamma}$0 and 1$^{\rm st}$ 
at the photon energy $1.7$~meV for the $x$-polarization is larger than 
that of the $y$-polarization. This is caused by the Rabi-splitting between 1${\gamma}$0 and 1$^{\rm st}$ for 
the $x$-polarization is larger than that of the $y$-polarization (see \fig{fig02}).
In contrast to the mentioned current peak, the broadening of the current peak formed at the photon energy $2.7$~meV is 
larger for the $y$-polarization. The reason is that the Rabi-splitting between 1${\gamma}$0 and 2$^{\rm nd}$ 
is larger for the $y$-polarization than the $x$-polarization. 
As we have mentioned before, the geometry of the states plays an essential role here. 
The first-excited state is more polarizable 
in the $x$-direction while the second-excited state by contrast is more polarizable in the $y$-direction.
We should mention that an intraband transition occurs between the aforementioned resonant states and it has a major role 
in the current transport~\cite{Nzar-arXiv_article_2019}. 
These intraband transitions can be tuned by other physical parameter of the system such as the electron-photon 
coupling strength and the cavity-photon reservoir coupling strength as is shown below.

\begin{figure}[htb]
  \includegraphics[width=0.5\textwidth,angle=0,bb=50 70 410 260]{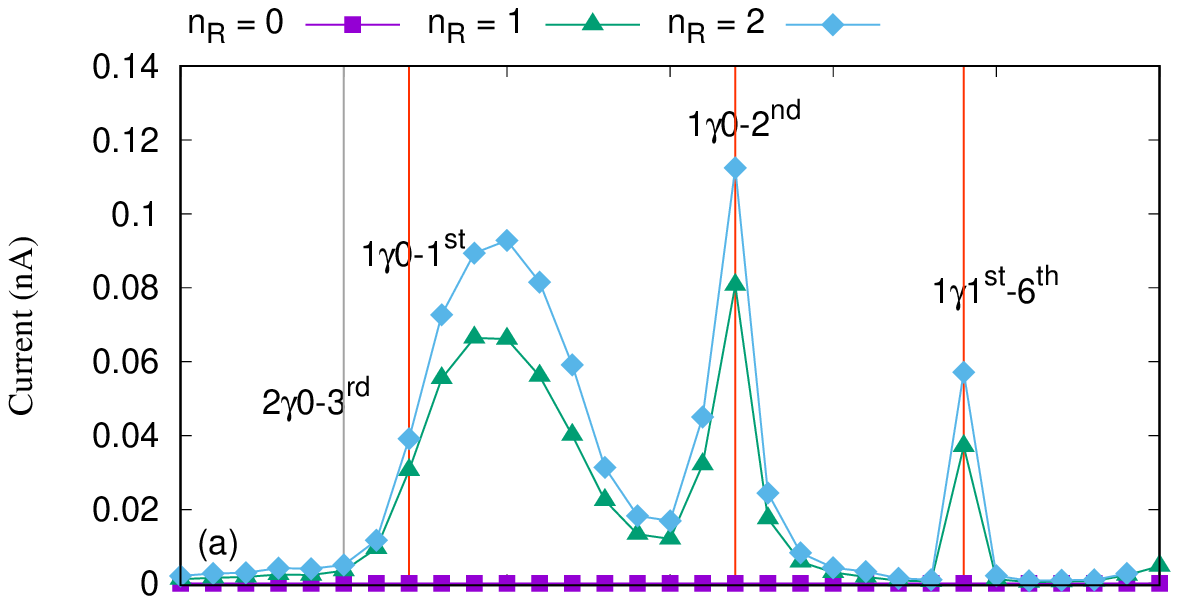}\\
  \includegraphics[width=0.5\textwidth,angle=0,bb=50 55 409 210]{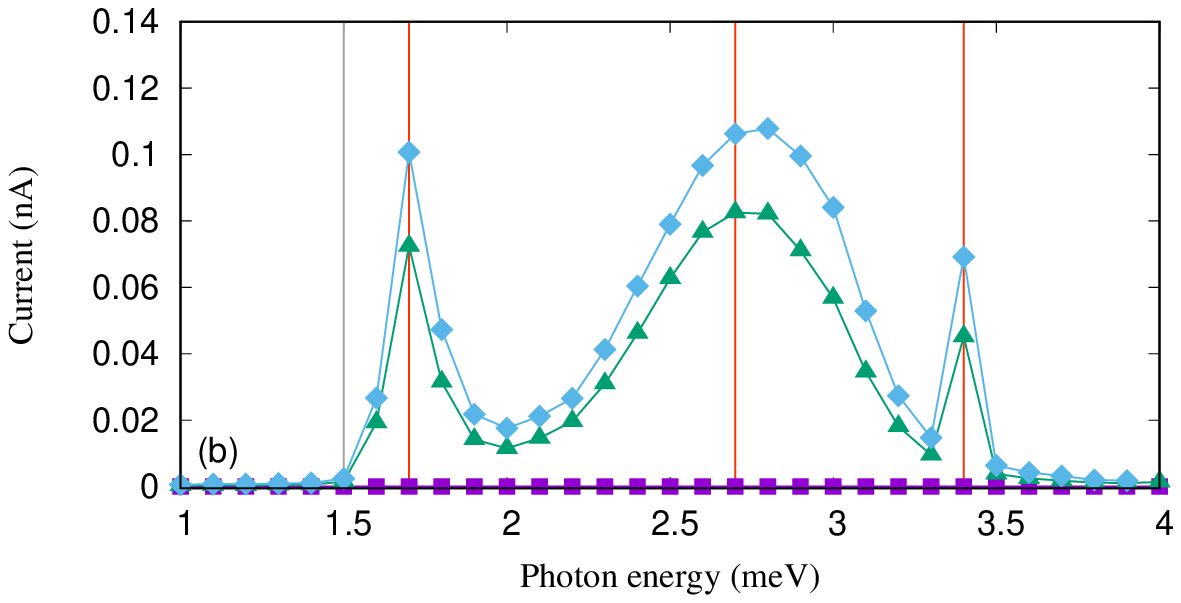}
       \caption{Current as a function of the photon energy for the photon number $n_\mathrm{R} = 0$ (purple squares), 
          $1$ (green triangles), and $2$ (blue diamonds)
          in the case of $x$- (a) and $y$-polarized (b) photon field. 
          The electron-photon coupling strength is $g_{\gamma} = 0.1$~meV, and $\kappa = 10^{-5}$ meV.
          The vertical red lines indicates the location of the main resonance states.
          The chemical potential of the left lead is $\mu_L = 1.65$~meV and the right lead is $\mu_R = 1.55$~meV.
          The magnetic field is $B = 0.1~{\rm T}$, $eV_{\rm g} = 0.651$~meV, $T_{\rm L, R} = 0.5$~K, and $\hbar \Omega_0 = 2.0~{\rm meV}$.}
\label{fig03}
\end{figure}

We now tune the photon number in the photon reservoir, $n_R$, and see it's influence 
on the current transport properties of the QD system in \fig{fig03}, where the cavity-reservoir coupling strength 
is assumed to be $10^{-5}$ meV 
and $g_{\gamma} = 0.1$~meV.
As the photon number is increased the participation of the photon replicas in the 
electron transport is enhanced. As a result, the current is slightly increased for
the case of two photons (green diamonds) for both photon polarizations. 
This happens as the photon
replicas are not pure simple perturbational states with
an integer number of photons, but instead contain states
with $0$, $1$, and $2$ photons at least to some amount.

It should be noted that if the mean number of photon is zero, $n_\textrm{R} = 0$, the current is very close to zero  
which is due to inactivated photon replica states in the transport in the absence of flow of photons 
into the cavity from the reservoir. In addition, the QD system is in a Coulomb blocking regime in the steady state when 
$n_\mathrm{R}=0$ and the charging of $1\gamma$0 and 1${\gamma}$1$^{\rm st}$ is thus approaching zero. 
These effects lead to a vanishing current. 

\begin{figure}[htb]
  \includegraphics[width=0.5\textwidth,angle=0,bb=50 70 415 260]{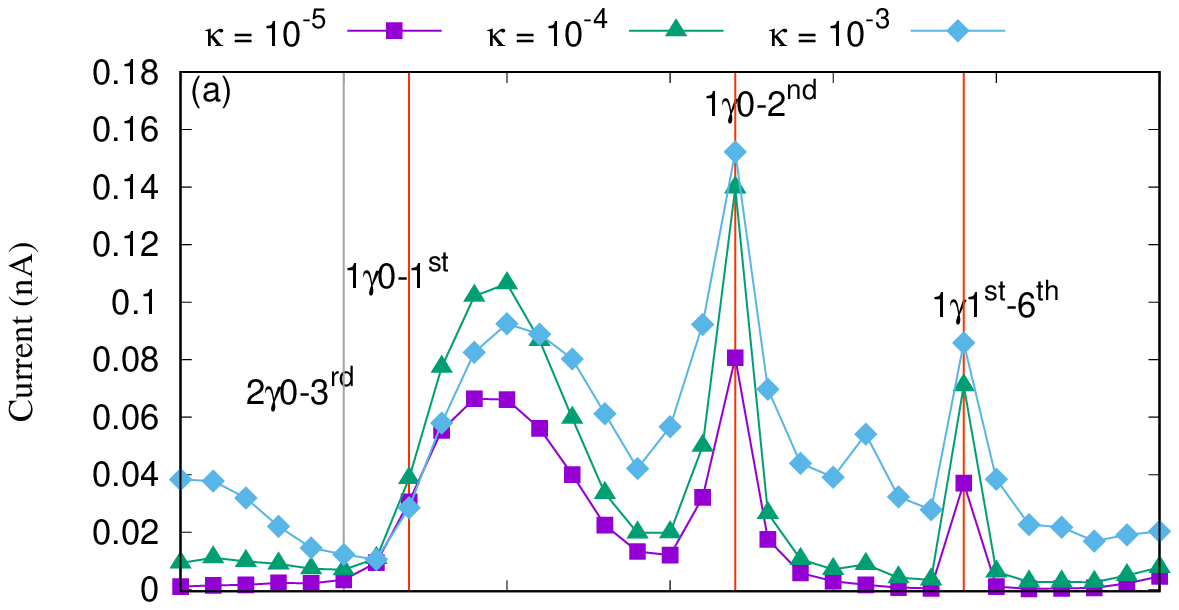}\\
  \includegraphics[width=0.5\textwidth,angle=0,bb=50 55 415 222]{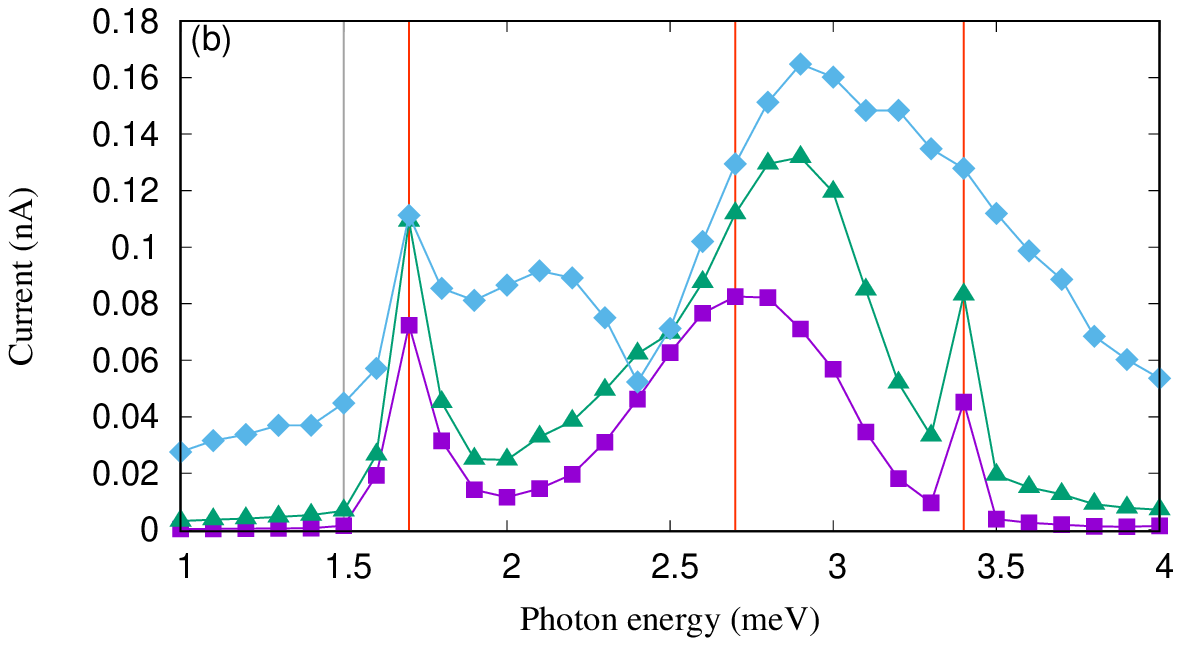}
       \caption{Current versus the photon energy for the cavity-reservoir coupling  $\kappa = 10^{-5}$ meV (purple squares), 
          $10^{-4}$ (green triangles), and $10^{-3}$ (blue diamonds)
          in the case of $x$- (a) and $y$-polarized (b) photon field. 
          The electron-photon coupling strength is $g_{\gamma} = 0.1$~meV, and $n_\textrm{R} = 1$.
          The vertical red lines indicates the location of the main resonance states.
          The chemical potential of the left lead is $\mu_L = 1.65$~meV and the right lead is $\mu_R = 1.55$~meV.
          The magnetic field is $B = 0.1~{\rm T}$, $eV_{\rm g} = 0.651$~meV, $T_{\rm L, R} = 0.5$~K, and $\hbar \Omega_0 = 2.0~{\rm meV}$.}
\label{fig04}
\end{figure}

We further investigate the transport characteristics by tuning the cavity-reservoir coupling and fix 
the photon number in the reservoir. Figure \ref{fig04} shows the current versus the photon energy for 
different values of the cavity-reservoir coupling strength in the case of $x$- (a) and $y$-polarized (b) photon field
where the electron-photon coupling strength is fixed at $g_{\gamma} = 0.1$~meV and the mean value of photons with
the particular energy in the reservoir is $n_\textrm{R} = 1$, respectively.
The current is enhanced with the cavity-reservoir coupling overall for both photon polarizations. 
This shows that the cavity-reservoir coupling influences the intraband transitions that occur between the resonance states 
forming the Rabi-resonant pairs. 

\begin{figure}[htb]
  \includegraphics[width=0.42\textwidth,angle=0,bb=50 70 410 260]{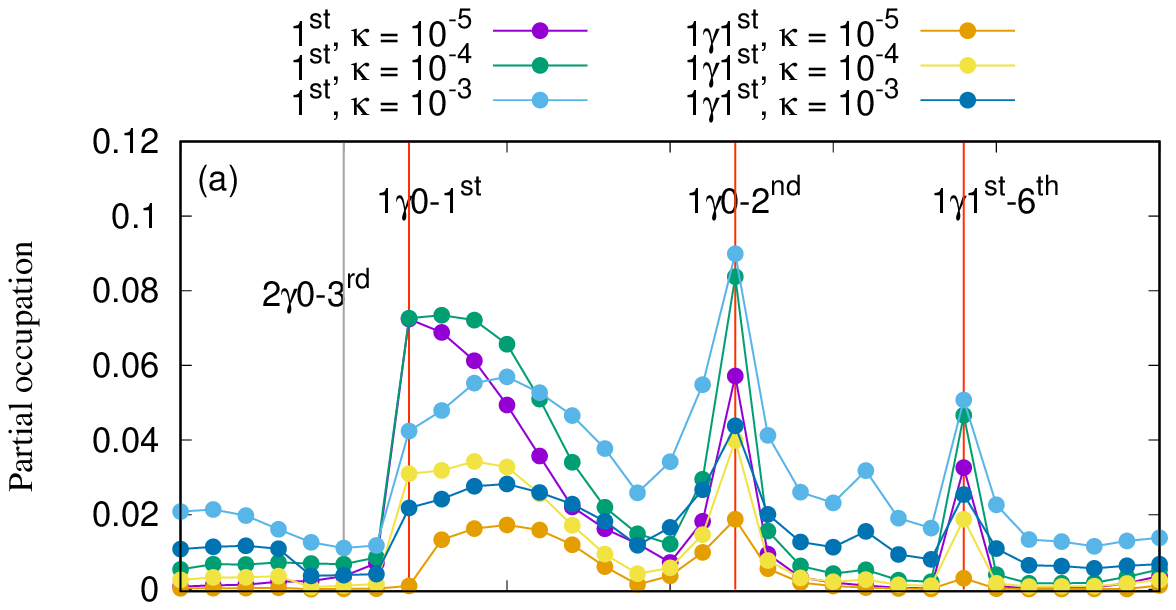}\\
  \includegraphics[width=0.42\textwidth,angle=0,bb=50 55 409 210]{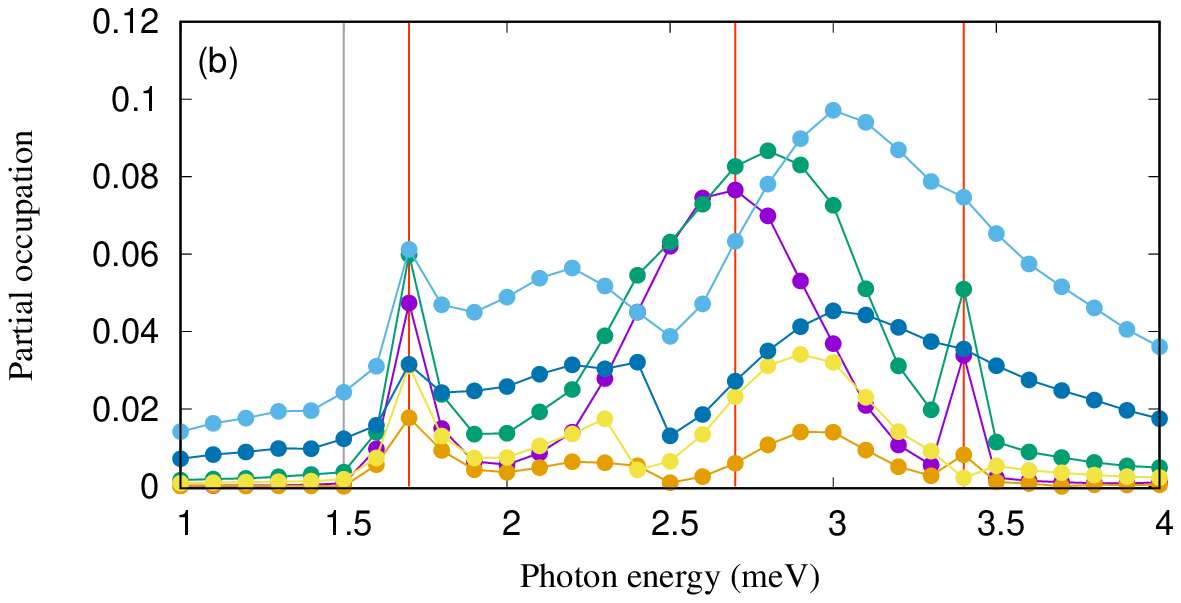}\\
  \includegraphics[width=0.42\textwidth,angle=0,bb=45 70 410 260]{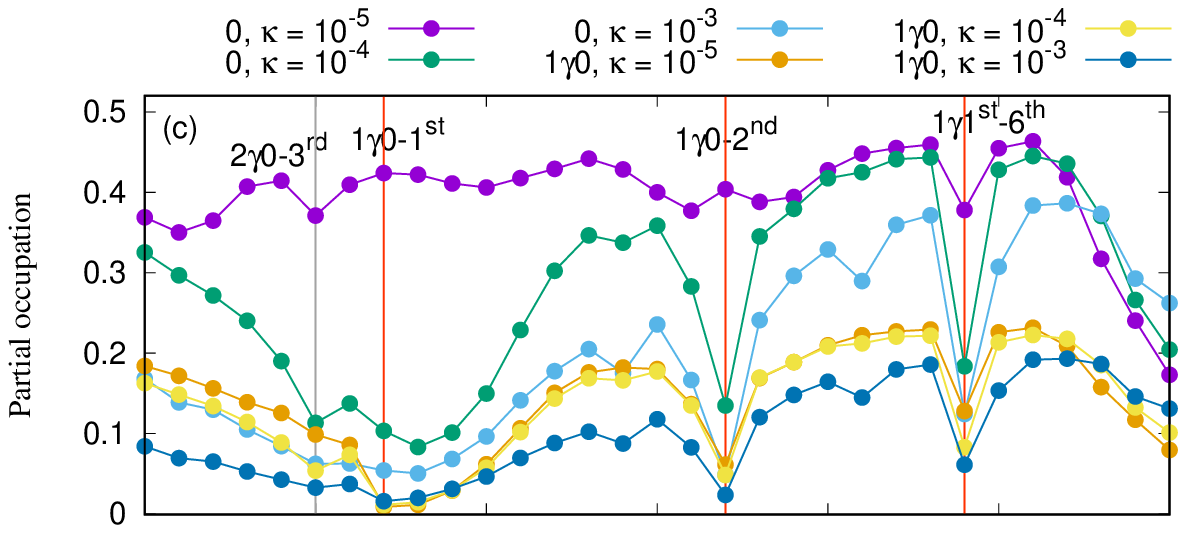}\\
  \includegraphics[width=0.42\textwidth,angle=0,bb=50 55 409 210]{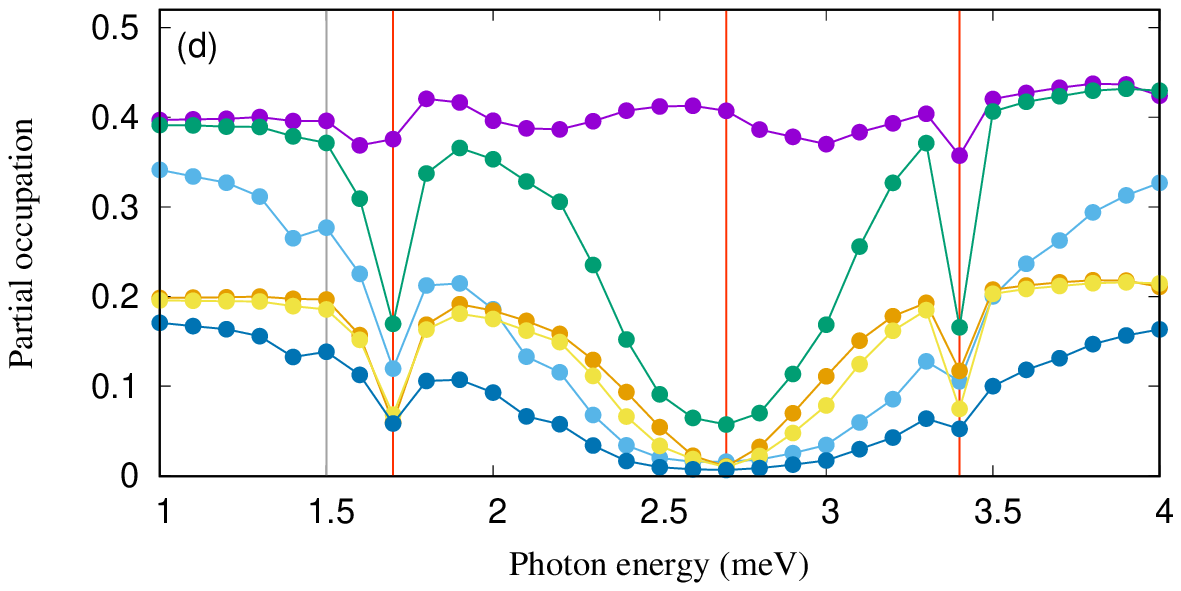}
       \caption{The partial occupation of the spin-up component of 1$^{\rm st}$ and 1${\gamma}$1$^{\rm st}$ 
       for different cavity-reservoir coupling
       is plotted $x$- (a) and $y$-polarized photon field (b). 
       Furthermore, the partial occupation of the spin-up component of 0 and 1${\gamma}$0 for different cavity-reservoir coupling  
       is presented for $x$- (c) and $y$-polarized photon field (d).
          The cavity-reservoir coupling is assumed to be $\kappa = 10^{-5}$ (purple for 0, and brown for 1$^{\rm st}$), 
          $10^{-4}$ (green for 0, and yellow for 1$^{\rm st}$), and $10^{-3}$ (light blue for 0, and dark blue for 1$^{\rm st}$)
          in the case of $x$- (a) and $y$-polarized (b) photon field. 
          The electron-photon coupling strength is $g_{\gamma} = 0.1$~meV, and $n_\textrm{R} = 1$.
          The vertical red lines indicates the location of the main resonance states.
          The chemical potential of the left lead is $\mu_L = 1.65$~meV and the right lead is $\mu_R = 1.55$~meV.
          The magnetic field is $B = 0.1~{\rm T}$, $eV_{\rm g} = 0.651$~meV, $T_{\rm L, R} = 0.5$~K, and $\hbar \Omega_0 = 2.0~{\rm meV}$. }
\label{fig05}
\end{figure}

In order to explain the current enhancement, we refer to the partial occupation of the most active states in the transport which are 
the first-excited state, 1$^{\rm st}$, and the first excitation thereof, 1${\gamma}$1$^{\rm st}$, in \fig{fig05} 
for the $x$- (a) and $y$-polarization  (b) on one hand, 
and on the other hand, the occupation of the ground-state, 0, and the first-excitation thereof, 1$\gamma$0, is presented in 
\fig{fig05} for the $x$- (c) and $y$-polarized photon field (d). We should mention that the \fig{fig05} shows only the spin-up component 
of the corresponding states, and the spin-down component is qualitatively the same (not shown). 
Increasing the coupling strength of cavity-photon reservoir, the occupation of the first-excited state and the first excitation
thereof is enhanced for both direction of the photon polarization while the occupation of the ground state and the excitation 
thereof is suppressed especially for the Rabi-resonant states. 
The first indication of charging of 1$^{\rm st}$ and 1${\gamma}$1$^{\rm st}$, and discharging of 0 and 1${\gamma}$0 for 
the Rabi-resonant states is a confirmation of the intraband transition occurring between the states.

Increasing the cavity-photon reservoir coupling strength, these intraband 
transitions become weak especially at $\kappa = 10^{-3}$. Therefore, the current going through $0$ and 1${\gamma}$0 is almost 
blocked but the current via 1$^{\rm st}$, and 1${\gamma}$1$^{\rm st}$ is increased which in turn increase the total current 
through the QD system because 1$^{\rm st}$ is confined in the bias window.
We are seeing here a manifestation of the Purcell effect, that was originally stated about the enhancement 
of radio wave emission of atoms in photon-cavities \cite{PhysRev.69.681}, but here it manifests itself in the enhanced current 
peaks of electrons through the cavity. 

The last test of our calculation is the influences of electron-photon coupling strength 
between the electrons in the quantum dot system and the photons in the cavity on the transport properties.
\begin{figure}[h]
\centering
    \includegraphics[width=0.23\textwidth,angle=0]{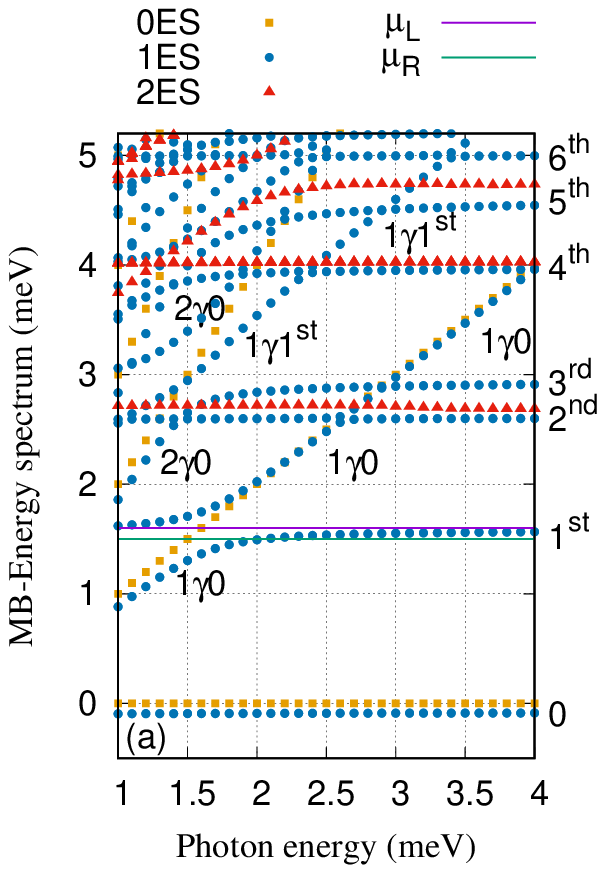}
    \includegraphics[width=0.23\textwidth,angle=0]{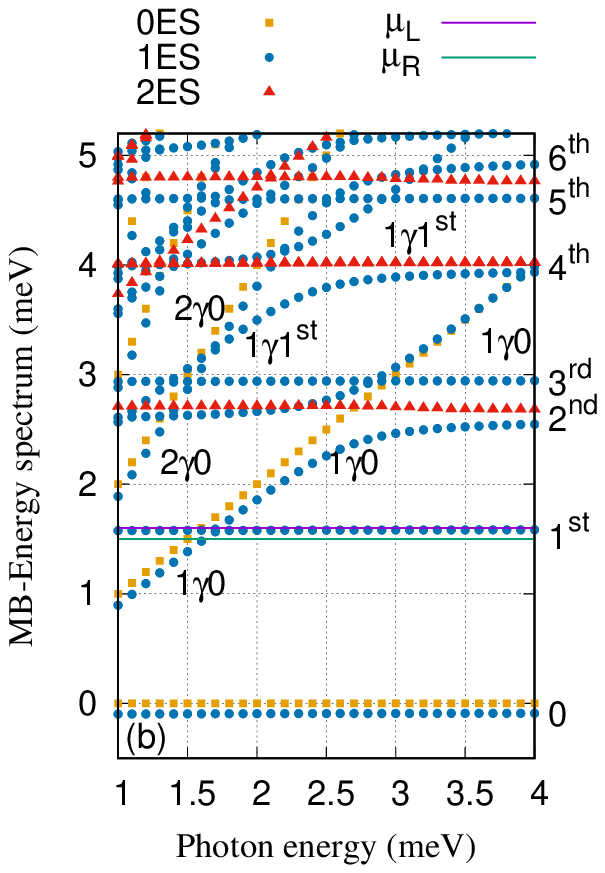}
 \caption{Many-Body energy spectra of the QD system coupled to the photon field as a function of 
        the photon energy for $x$- (a) and $y$-polarized (b) photon field,
        where 0ES (brown squares) are zero-electron states, 1ES (blue circles) are one-electron states, 
        and 2ES (red triangles) are two-electron states.
        The chemical potential of the left lead is $\mu_L = 1.65$~meV (purple line) and the right lead is $\mu_R = 1.55$~meV (green line).
        0 indicates the one-electron ground-state energy, $1\gamma$0 and $2\gamma$0 refer to the one- and two-photon replica
        of the 0, and 1$^{\rm st}$, 2$^{\rm nd}$, 3$^{\rm rd}$, 4$^{\rm th}$, 5$^{\rm th}$, 6$^{\rm th}$ display the one-electron first-, second-, third-, fourth-, fifth- and sixth-excited state, 
        respectively. The $1\gamma$1$^{\rm st}$ indicates the one-photon replica state of the 1$^{\rm st}$.
        The electron-photon coupling strength $g_{\gamma} = 0.3$~meV, the cavity-reservoir coupling  $\kappa = 10^{-5}$ meV, and 
        the photon number initially in the cavity $n_\mathrm{R} = 1$.
        The magnetic field is $B = 0.1~{\rm T}$, $eV_{\rm g} = 0.651$~meV, $T_{\rm L, R} = 0.5$~K and $\hbar \Omega_0 = 2.0~{\rm meV}$.}
\label{fig06}
\end{figure}
Figure \ref{fig06} displays the many-body energy spectrum of the QD system coupled to the cavity 
for both $x$- (a) and $y$-polarized (b) cavity-photon field where the electron-photon coupling strength is tuned to 
$g_{\gamma} = 0.3$~meV. Comparing to the energy spectrum presented in \fig{fig02}, where the electron-photon coupling 
strength is weaker, $g_{\gamma} = 0.1$~meV, some changes in the energy spectrum can 
be seen~\cite{Arnold_2014,Nzar-arXiv_article_2019_1}. 
For instance, the Rabi-splitting between 1$^{\rm st}$ and 1${\gamma}$0 at the photon energy $1.7$~meV becomes larger here for the 
$x$-polarized photon field (see \fig{fig06}a). Furthermore, the Rabi-splitting between  1${\gamma}$0 and 2$^{\rm nd}$ at the photon energy $2.7$~meV here
is much larger for the $y$-polarization comparing to the case when $g_{\gamma} = 0.1$~meV (see \fig{fig06}b).
The first-photon replica state, 1${\gamma}$1$^{\rm st}$, is not resonant with the sixth-excited state, $6^{\rm th}$, 
anymore here while a strong Rabi-splitting between these two state was seen at $g_{\gamma} = 0.1$~meV (see \fig{fig02}) 
especially for the $y$-polarization. 

\begin{figure}[htb]
    \includegraphics[width=0.47\textwidth,angle=0,bb=70 70 410 260]{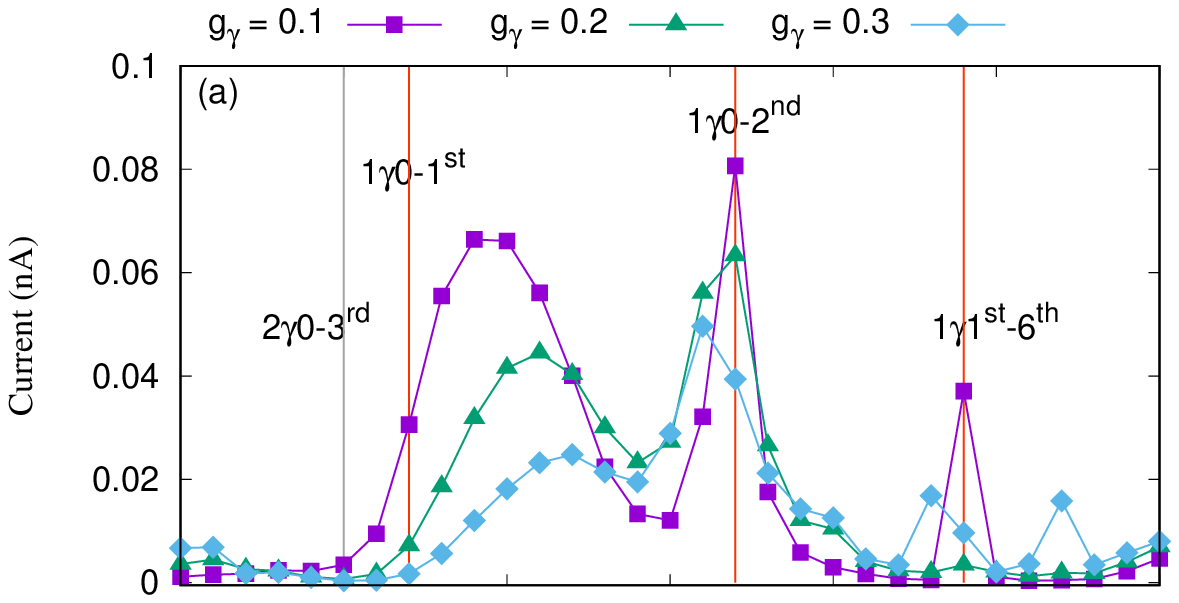}\\
    \includegraphics[width=0.47\textwidth,angle=0,bb=70 55 410 235]{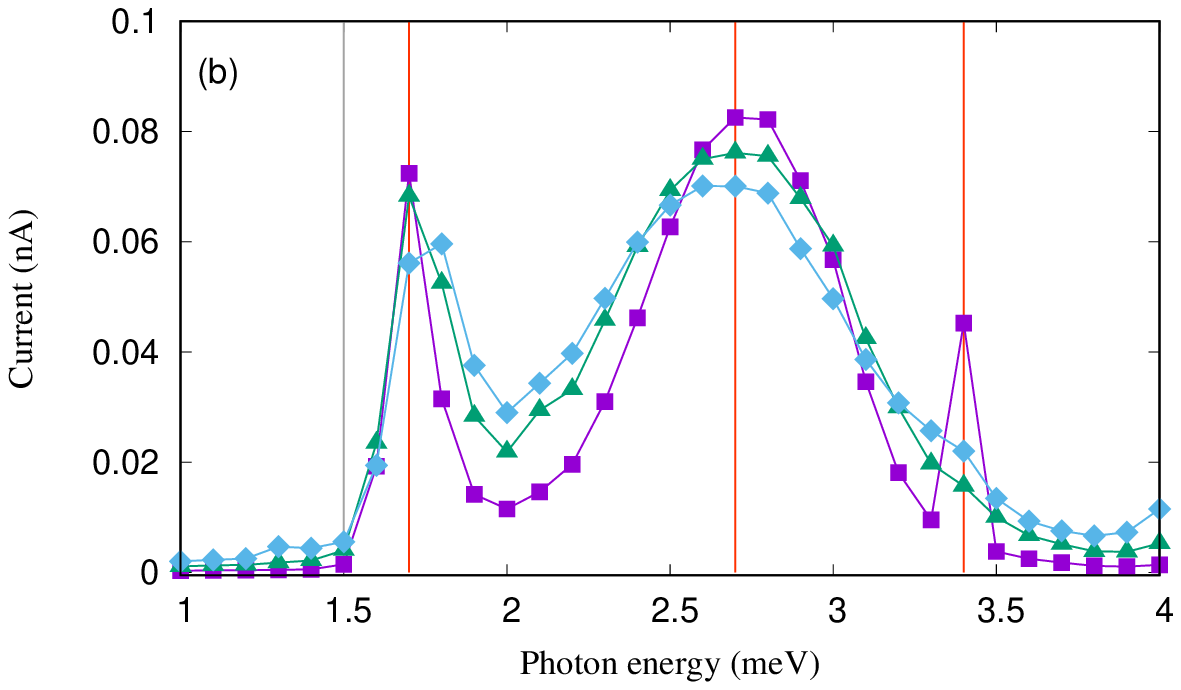}
       \caption{Current as a function of the photon energy for the electron-photon coupling strength 
          $g_{\gamma} = 0.1$  (purple squares), 
          $0.2$ (green triangles), and $0.3$~meV (blue diamonds)
          in the case of $x$- (a) and $y$-polarized (b) photon field. 
          The cavity-reservoir coupling  is $\kappa = 10^{-5}$, and $n_\textrm{R} = 1$.
          The vertical red lines indicates the location of the main resonance states.
          The chemical potential of the left lead is $\mu_L = 1.65$~meV and the right lead is $\mu_R = 1.55$~meV.
          The magnetic field is $B = 0.1~{\rm T}$, $eV_{\rm g} = 0.651$~meV, $T_{\rm L, R} = 0.5$~K, and $\hbar \Omega_0 = 2.0~{\rm meV}$.  }
\label{fig07}
\end{figure}

The current as a function of the photon energy for three values of the electron-photon coupling strength is shown 
in \fig{fig07} for the $x$- (a) and $y$-polarized (b) photon field. 
The current decreases with increasing 
electron-photon coupling strength for both direction of photon polarization. We start with the case of $x$-polarization 
(see \fig{fig07}a), the current suppression in the leftmost peak is observed at high $g_{\gamma} =  0.3$~meV (blue squares) 
which is due to the larger Rabi-splitting between 1$^{\rm st}$ and 1${\gamma}$0 for the photon energy $1.7$~meV. 
The Rabi oscillation between these two states is thus increased and 
in turn the current is diminished. Furthermore, the positions of the two other peaks are shifted at $g_{\gamma} =  0.3$~meV 
since the locations of Rabi-splitting forming the two peaks are moved.

For the $y$-polarized photon field (\fig{fig07}b), the current of the leftmost  peak is slightly changed with electron-photon 
coupling strength because the Rabi-splitting of the corresponding states is not much influenced by the photon polarization as 
is shown in \fig{fig02}b and \fig{fig06}b. 
In addition, the broadening of the current peak formed due to the Rabi-splitting between 1${\gamma}$0 and 2$^{\rm nd}$  
at $2.7$~meV is increased at higher electron-photon coupling strength. The last current peak at the photon energy $3.4$~meV 
vanishes since the anti-crossing between 1${\gamma}$1$^{\rm st}$ and $6^{\rm th}$ is not found anymore at 
$g_{\gamma} =  0.3$~meV  (see \fig{fig06}b).

\section{Summary}\label{Sec:Conclusion}

To summarize our results, we have shown that the photon polarization, 
the electron-photon coupling strength, the coupling strength of cavity-photon reservoir, 
and the mean photon number in the environment/reservoir can be used to control the resonance current peaks emerging 
due to the Rabi-resonant states of a quantum dot system coupled to a photon cavity and an external photon reservoir. 
We show that the photon polarization and the electron-photon coupling strength 
play an important role in the forming of Rabi-resonant states which in turn generate resonant current peaks.
Furthermore, increasing the cavity photon coupling to the environment, $\kappa$, opening for faster flow of
photons into and out of the cavity, the photon replica states are further activated leading 
to enhancement the electron transport. This phenomena demonstrates the Purcell effect \cite{PhysRev.69.681} 
observed through current transport. 
Finally, by tuning the cavity-photon coupling strength the intraband transition between 
the Rabi-resonant states can be controlled.

\bigskip

\begin{acknowledgments}
This work was financially supported by the Research Fund of the University of Iceland,
the Icelandic Research Fund, grant no.\ 163082-051, 
and the Icelandic Infrastructure Fund. 
The computations were performed on resources provided by the Icelandic 
High Performance Computing Center at the University of Iceland.
NRA acknowledges support from University of Sulaimani and 
Komar University of Science and Technology.
\end{acknowledgments}

%

%

%
\end{document}